%% file: SCCurrent_5.tex
\documentclass[aps,prl,twocolumn,superscriptaddress]{revtex4-1}

\usepackage{graphicx}  
\usepackage{amsmath} 

\usepackage{graphicx}
\usepackage{amsmath,amssymb,amsfonts}
\usepackage{textcomp}
\usepackage{hyperref}
\usepackage{gensymb}
\usepackage{verbatim}
\usepackage{braket}

\hypersetup{breaklinks=true,colorlinks=true,urlcolor=black}
\usepackage{color}
\usepackage{soul}
\usepackage{gensymb}

\DeclareGraphicsExtensions{.jpg,.pdf,.png,.eps}

\begin{document}


\title{Geometric origin of supercurrents in Berry phase: Formula for computing currents from wavefunctions with correlation and particle number variation}



\author{B.Q.~Song}
\affiliation{Ames National Laboratory, Iowa State University, Ames, Iowa 50011, USA}
\affiliation{Department of Physics and Astronomy, Iowa State University, Ames, Iowa 50011, USA}
\affiliation{Department of Physics, University of Houston, Houston, Texas 77204, USA}
\affiliation{Texas Center for Superconductivity, University of Houston, Houston, Texas 77204, USA}

\author{J.D.H.~Smith}
\affiliation{Ames National Laboratory, Iowa State University, Ames, Iowa 50011, USA}
\affiliation{Department of Mathematics, Iowa State University, Ames, Iowa 50011, USA}
\author{J.~Wang}
\affiliation{Ames National Laboratory, Iowa State University, Ames, Iowa 50011, USA}
\affiliation{Department of Physics and Astronomy, Iowa State University, Ames, Iowa 50011, USA}


\date{\today}

\begin{abstract}
The complexity of itinerant and many-body nature in Bardeen–Cooper–Schrieffer (BCS) wavefunctions has traditionally led to the use of coarse-grained order parameters for describing currents in superconductors (SC), rather than directly utilizing wavefunctions. In this work, we introduce a phase-based formula that enables the direct computation of currents from microscopic wavefunctions, accounting for correlation and particle number variations. Interestingly, the formulation draws parallels with insulators, suggesting a unified framework for understanding (intra-band) charge transport across two extremes of conductivity. A “group velocity” current $J_{band}{\propto}\frac{1}{\hbar}{\partial}_kE(k)$ is derived from Berry phase, independent of wave package dynamics, robust against correlation. Additionally, we identify a correlation-driven contribution, $J_{corr}$,  which reveals that the pairing correlations ${\langle}c_kc_{-k}{\rangle}$ among “dancing partners” provide a current component beyond the velocity operator.
\end{abstract}



\maketitle

\section{I. Introduction}
Over the past few decades, geometric concepts -- such as Berry phase \cite{c1,22,c2,24}, quantum geometry \cite{c3,c4,c5,23,27,58,62,67,c6,c7,c8} -- have played central roles in deepening our understanding of condensed matter systems. They emphasize the structure of eigenvectors, instead of eigenvalues. That is, certain observables \cite{58,61,c3,c9,c10,c11,c12} are insensitive to the spectrum but closely related to “geometric quantities” such as connection \cite{c2,27,25,26,c13}, curvature \cite{c10,c11,c14,c15}, metric \cite{c3,c5,23,c8,58}, measure \cite{63,c16,c17} – these quantities recover the geometric intuitions in abstract spaces, including tangent vectors, curved surfaces, distances, volumes, bringing up novel notions and tunability \cite{c3,c4,c5,23,58,1a,c18,c19}. 

An area tied to the theme is the emerging role of quantum geometry/topology in superconductivity \cite{c4,c5,c12,23,58}, bringing together a younger field with a recurring subject since the early 20th century. For instance, quantum metric \cite{c5,23} influences the superfluids formed in the ground state (GS), especially in scenarios of flat-bands \cite{c5,58,67,c20}.

This work uncovers a formula that links the current density $J$ in superconductors (SCs) to the Berry phase, which is defined by the geometry of wave functions. Unlike previous protocols \cite{c4,c5,23,67} in which geometry enters at the level of characterizing a particular state (e.g., the superconducting GS, with geometry entering through the free energy of that state \cite{c5}), geometry appears at the level of the generic position operator $\hat{r}$. More concretely, Berry phase enters the general procedure of extracting expectation value ${\langle}r{\rangle}$, not limited by a specific state $\langle..\rangle$. Thus, the present result can be viewed as a case study of evaluating $\langle\dot{r}\rangle$ for BCS wave functions, thanks to its analytic forms and well-understood properties.

The Berry-phase current was earlier invented for insulators \cite{25,26}, and this work extends the formulation from insulators to SC. Consequently, the currents in the SC and insulator can be similarly calculated and understood, despite their extremely different conductivity.

This unified current formulation takes the microscopic electronic wavefunction as its fundamental input, thereby placing the pair-correlated BCS state on equal footing with non-SC states or other forms of correlation. In other words, it does not distinguish between different types of carriers, whether they are single quasi-particles or Cooper pairs.

The geometry-based formula can re-derive the conventional forms of current (eigenvalue slopes) without invoking ``group velocity" \cite{19,20} or other semi-classical notions. Furthermore, it reveals an additional term, previously overlooked, that emerges when inversion symmetry (IS) is broken. This highlights the importance of investigating non-reciprocal superconductivity \cite{45,46,c21,c22}. 

Since the BCS wavefunction has long been known \cite{1,2}, one may regard it as already fully understood. However, the history of noninteracting electron bands offers a cautionary example: despite their exact solvability and apparent simplicity \cite{c23}, crucial features were initially overlooked, most notably the topological structure of the wave function \cite{c12} and its profound consequences for edge and surface phenomena, such as the quantum Hall effect (QHE) \cite{c10,c11}. 

This raises a major question. Even when the wavefunction is known, have \textit{all} physical observables and consequences been extracted? For QHE, the edge observables were overlooked \cite{c10}, and their subsequent recognition gave rise to topological physics \cite{c23,c12,c11}. Likewise, for SC, we ask whether $J$ is completely yielded from the BCS wave function? Whether a fuller understanding reveals unexpected connection with other areas of physics?


Thus, the paper does not merely discuss an ``internal" question of SC, but one arises when SC is connected to non-SC (e.g., insulators), ultimately shedding new light back onto SC. On the other hand, the present discussion does not involve the tiers, such as pairing mechanism, finite temperature. Its main limitation is that the formula requires knowledge of the microscopic wave function, which may be inaccessible, especially in unconventional SC \cite{47,48,49,64,65,66}.


Section 2 explains how extending the Berry-phase formulation of current from insulator to SC resolves the divergent $\langle r \rangle$ issue and avoids the approximations inherent in the velocity operator $\hat{v}$ method for defining currents. Section 3 presents the generalization and shows that $J:=\rho\langle \dot{r} \rangle$ not only reproduces the current associated with $\hat{v}$ but also yields an additional $J_{corr}$ that purely arises from correlation change. This term exists when IS is broken, and experimental detection is discussed. Section 4 discusses the physical implications and its connections to other areas of physics.




\section{II. Existing (intra-band) formula: Velocity operator \& Phase-based current}
For a non-local state, such as Bloch or BCS states, the integral $\langle r \rangle:=\int_{\infty}\varphi^*r\varphi dr$ is divergent, which prevents a direct definition as $J\sim\langle\dot{r}\rangle$ and necessitates alternative approaches to extract observable ``current" from the wavefunction. A common method (for intra-band) is velocity operator $\hat{v}$ \cite{7,8}, 
\begin{equation}
\begin{split}
j(r)={\Re}[{\varphi}^*q\hat{v}{\varphi}]={\Re}[{\varphi}^*\frac{q}{m}(-i{\hbar}{\nabla}_r-q\textbf{A}){\varphi}].
\end{split}
\label{eq1}
\end{equation}
Equation (\ref{eq1}) outputs a field $j(r)$ (or Fourier transform $j(k)$) \cite{11,12,13} rather than an expectation value $J$. A crucial difference is that standard quantum expectation involves ${\int}_{\infty}dr$, but Eq. (\ref{eq1}) does not. Consequently, $v(r)$ is not guaranteed to be real-valued, and $\Re$ must be imposed. In concept, $j(r)$ assigns a definite momentum (via $\hat{v}$) to a definite position, violating uncertainty principle, thus $r$  is interpreted as a coarse-grained spot. In context of SC, $j(r)$ gives a semi-classical distribution of currents with a resolution of the coherence length. 

Thus, the locality of $\hat{v}$ inevitably involves semi-classical approximations. For example, applying it to Bloch waves $\frac{\hbar}{m}{\int}{\langle}\hat{v}{\rangle}dr$ leads to the ``group velocity" $\frac{1}{\hbar}{\partial}_kE(k)$ of a wave packet \cite{19,20}. Equation (\ref{eq1}) respects probability conservation $\dot{\rho}+{\nabla}{\cdot}j(r)=0$ thus provides a plausible local forms of currents, commonly useful in expressing free energy density. However, in view of the divergence of $\hat{v}$ for neutral spinful particles ($q=0, s{\neq}0$) \cite{10}, such as neutron (due to extra term ${\sim}1/qs$) \cite{7}, non-local contribution might be indispensable. 

Another approximation peculiar for SC is that $\varphi$ is Ginzburg-Landau order parameter (OP) \cite{5,14} with effective charge and mass: $q=-2e$, $m=2m_e$, rather than the many-body electronic wavefunction $\varphi(r_1,r_2...)$. The coordinate $r$ in $\varphi(r)$, $\textbf{A}(r)$, etc., is a general spatial position, not for a specific electron. It is unclear how to adopt $\hat{v}$ for multiple coordinates $r_1,r_2...$. For example, it is unknown how to modify vector potential $\textbf{A}(r)$ for higher dimensions in respecting gauge \cite{16,17,18}, not speaking of how to handle infinite particles or vacuum. (Note that $\hat{v}$ is void for a vacuum state, however, shortly we will see the “nothing” state still gives subtle influence on $J$ given it enters a superposition.)


Using OP leads to fragmented current formulations that depend on carrier species. For example, OP of SC is tied to the Cooper pair rather than electron carriers. Then, the separation between supercurrent $j_s$ and normal current $j_n$ emerges naturally, as in London equations \cite{1,5}. Ideally, however, a unified formulation should eliminate such a division, reflecting the common physical origin of ``charge movement", and the fact that $j_s+j_n$ together constitute the current in Maxwell’s equations \cite{1,2,5}.

In view of these limitations, improvements could be achieved by (i) removing the semi-classical approximation (e.g., group velocity of wave packet) and the assumption of locality; (ii) replacing the OP by the microscopic electronic wave function as the input for calculating $J$; (iii) restoring the original definition of current based on ${\langle}\dot{r}{\rangle}$. 

Progress is inspired by \textit{phase-based formulation of current} in modern theory of polarization \cite{24,25,26}.
\begin{equation}
\begin{split}
J={\rho}{\langle}\dot{r}{\rangle}={\rho}\frac{a}{2\pi}\dot{\vartheta}=\frac{q}{V_{\text{cell}}}\frac{a}{2\pi}{\partial}_t{\oint}{\langle}A_{n,k}|i{\partial}_kA_{n,k}{\rangle}dk,
\end{split}
\label{eq2}
\end{equation}
where position ${\langle}r{\rangle}$ is linked to Berry phase ${\vartheta}$. The Berry phase $\vartheta$ is evaluated by Berry connection ${\langle}A_{n,k}|i{\partial}_kA_{n,k}{\rangle}$ defined for the periodic part $|A_{n,k}{\rangle}$ of the $n^{th}$ Bloch band \cite{24,27}. 

Equation~(\ref{eq2}) is a non-local formulation; its global nature -- integration over Brillouin zone (BZ) -- naturally respects gauge symmetry \cite{29}. It is purely quantum, independent of the semi-classical picture. Its input is a microscopic wavefunction, rather an OP field. It provides a convergent ${\langle}r{\rangle}$ \cite{28}, allowing one to define currents with ${\langle}\dot{r}{\rangle}$. It has proved effective in calculating polarization and adiabatic currents \cite{24,26}. 

The appealing features underscore the Berry phase as a promising candidate for formulating currents. On the other hand, one may ask how broadly Eq.~(\ref{eq2}) can be applied? Unfortunately, it is presently limited to single-particle Bloch waves and mainly applies for intra-band transport of insulators. 

It is easy to show the difficulty for BCS state $|{\psi}_s{\rangle}$.
\begin{equation}
\begin{split}
|{\psi}_s{\rangle}={\prod}_j^N(u_{k_j}+v_{k_j}b_{k_j}^{\dagger})|{\Phi}_0{\rangle},
\end{split}
\label{eq3}
\end{equation}
where $u_{k_j}$, $v_{k_j}$ are coefficients, and $b_{k_j}^{\dagger}:=c_{k_j,{\uparrow}}^{\dagger}c_{-k_j,{\downarrow}}^{\dagger}$, two Fermion (creation) operators for a time reversed pairs acting on vacuum $|{\Phi}_0{\rangle}$. Clearly, $|A_{n,k}{\rangle}$ in Eq.~(\ref{eq2}) cannot be plugged in with $|{\psi}_s{\rangle}$, first of all, due to the mismatch between their coordinates: $k$ for position operator $\hat{r}$ and multiple ${\lbrace}k_j{\rbrace}_N$ for $|{\psi}_s{\rangle}$. 

\section{III. Berry-phase formulation of current for superconductor}
Therefore, a crucial step is extending the variables in the formulation. Once the applicability of geometric formula Eq.~(\ref{eq2}) is generalized, its appealing features will be inherited by SC, and the problems related to fragmented formulations, semi-classical approximation, will be mitigated. 

To extend the variables, we make the following generalizations. First, generalize $\hat{r}$ with
\begin{equation}
\begin{split}
\hat{r}{\rightarrow}\frac{1}{N}{\sum}_j^Ni{\partial}_{k_j},~k{\rightarrow}{\sum}_j^Nk_j.
\end{split}
\label{eq4}
\end{equation}
Evidently, Eq.~(\ref{eq4}) satisfies $[\hat{r},k]=i$. Since the forms of $\hat{r}$ and $k$ are “solutions” for the commutator equation $[\hat{r},k]=i$, just like the familiar substitution $\hat{r}{\rightarrow}i{\partial}_k$ is, Eq.~(\ref{eq4}) serves as general forms of $\hat{r}$ and $k$ operators; the conventional case corresponds to $N=1$. Now, $\hat{r}$ can act on $|{\psi}_s{\rangle}$ with matched coordinates.

Second, re-write Eq.~(\ref{eq3}) in a convenient from of a $2^N$-dimensional vector
\begin{equation}
\begin{split}
|{\psi}_s{\rangle}=\begin{pmatrix} u_{k_1} \\ v_{k_1} \end{pmatrix}{\otimes}\begin{pmatrix} u_{k_2} \\ v_{k_2} \end{pmatrix}{\otimes}...\begin{pmatrix} u_{k_N} \\ v_{k_N} \end{pmatrix}
\end{split}
\label{eq5}
\end{equation}
where $u_k^2=1-v_k^2=\frac{1}{2}(1+\frac{{\epsilon}_k}{\sqrt{{\epsilon}_k^2+{\Delta}_k^2}})$, and ${\epsilon}_k$ is the band energy in the normal state. A parabolic approximation is usually taken: $\frac{{\hbar}^2k^2}{2m}-{\epsilon}_F$. ${\Delta}_k={\sum}_{k'}^NV_{k,k'}u_{k'}v_{k'}$ is the standard mean-field parameter \cite{5}, and $V_{k,k'}$ is attractive pairing potential. The Eq.~(\ref{eq3}) and Eq.~(\ref{eq5}) are required to preserve the inner product, namely isomorphism \cite{30}
\begin{equation}
\begin{split}
&{\prod}_j^N{\langle}{\Phi}_0|(u_{k_j}^{'*}+v_{k_j}^{'*}b_{k_j}){\cdot}{\prod}_l^N(u_{k_l}+v_{k_l}b_{k_l}^{\dagger})|{\Phi}_0{\rangle}\\
&={\prod}_j^N\begin{pmatrix} u_{k_j}^{'*} & v_{k_j}^{'*} \end{pmatrix}\begin{pmatrix} u_{k_j} \\ v_{k_j} \end{pmatrix}
\end{split}
\label{eq6}
\end{equation}
Thus, Eq.~(\ref{eq3}) and Eq.~(\ref{eq5}) are equivalent for a shared structure of space. Eq.~(\ref{eq5}) happens to be the \textit{Anderson pseudo-spin} \cite{16,31}, which is peculiar for BCS states. For a general correlated state, it requires isomorphism to determine the counterpart of Eq.~(\ref{eq5}).

Third, generalize the domains of integration. Remember Eq.~(\ref{eq2}) is “global” because ${\langle}r{\rangle}$ depends on the entirety of all Bloch waves over BZ, while ${\langle}r{\rangle}$ is diverging if only a single Bloch wave at local $k$ is counted. Now with coordinates ${\lbrace}k_j{\rbrace}_N$, a natural generalization is
\begin{equation}
\begin{split}
\frac{a}{2\pi}{\int}_0^{\frac{2\pi}{a}}dk~{\rightarrow}~\frac{a^N}{(2\pi)^{N}}{\int}_0^{\frac{2\pi}{a}}...{\int}_0^{\frac{2\pi}{a}}dk_1{\cdots}dk_N
\end{split}
\label{eq7}
\end{equation}
The parameter $N$ corresponds to the number of unit cells (i.e., the sites in the chain), \textit{not} the number of electrons. Thus, the validity of the current formulation is unaffected by whether the particle number is not a good quantum number.

Note that the generalized definition of $\hat{r}$ directly arises from the solutions Eq.~(\ref{eq4}) of commutator $[\hat{r},k]=i$, making no reference to details of the Hamiltonian or wavefunction, such as if the particle number is conserved. This is aligned with spirit of defining a quantum operator, for instance, the spin operator, which is merely derived with commutator $[s_i,s_j]=i{\hbar}{\epsilon}_{ijk}s_k$ independent of the form of the Hamiltonian. Thus, Eq.~(\ref{eq4}) provides a fundamental definition for current by $\langle\dot{r}\rangle$.

With Eq.~(\ref{eq4}),(\ref{eq5}),(\ref{eq7}), we find the position center ${\langle}r{\rangle}_s$ of a BCS state using
\begin{equation}
\begin{split}
&\frac{a^N}{(2\pi)^N}{\oint}_{{\lbrace}k_i{\rbrace}}\left({\bigotimes}_{l=1}^N\begin{pmatrix} u_{k_l}^* & v_{k_l}^* \end{pmatrix}\right)\times
\\&\rule{20mm}{0mm}\times
\left(\frac{1}{N}{\sum}_j^Ni{\partial}_{k_j}\right)\left({\bigotimes}_{l=1}^N\begin{pmatrix} u_{k_l} \\ v_{k_l} \end{pmatrix}\right)\\
&=\frac{a}{2\pi}\frac{1}{N}{\sum}_{j=1}^N{\int}_0^{\frac{2\pi}{a}}(u_{k_j}^*i{\partial}_{k_j}u_{k_j}+v_{k_j}^*i{\partial}_{k_j}v_{k_j})dk_j.
\end{split}
\label{eq8}
\end{equation}
By introducing
\begin{equation}
\begin{split}
|A_k{\rangle}:=\begin{pmatrix} u_k \\ v_k \end{pmatrix},~{\langle}A_k|:=\begin{pmatrix} u_k^* & v_k^* \end{pmatrix},
\end{split}
\label{eq9}
\end{equation}
we may express Eq.~(\ref{eq8}). as
\begin{equation}
\begin{split}
{\langle}r{\rangle}_s=\frac{a}{2\pi}{\int}_0^{\frac{2\pi}{a}}{\langle}A_k|i{\partial}_kA_k{\rangle}dk.
\end{split}
\label{eq10}
\end{equation}
The position center ${\langle}r{\rangle}_s$ for a BCS SC is formally identical with Eq.~(\ref{eq2}) for an insulator. Although $|A_k{\rangle}$ in Eq.~(\ref{eq9}) resembles a two-band state, it actually represents correlation within a single band. In general, $|A_k{\rangle}$ could be of any dimensions, relying on forms of correlation. However, the phase-based formulation Eq.~(\ref{eq10}) is generic. 

Then, we determine the time derivative ${\langle}\dot{r}{\rangle}_s$. Without loss of generality for intra-band, the correlation functions $u_k$ and $v_k$ are assumed time dependent.
\begin{equation}
\begin{split}
|{\psi}_s(t){\rangle}=e^{-iE(k_1,...k_N)t/\hbar}{\bigotimes}_{j=1}^N\begin{pmatrix} u_{k_j}(t) \\ v_{k_j}(t) \end{pmatrix},
\end{split}
\label{eq11}
\end{equation}
where
\begin{equation}
\begin{split}
E(k_1,...k_N):={\langle}{\psi}_s|H(k_1,...k_N)|{\psi}_s{\rangle}.
\end{split}
\label{eq12}
\end{equation}
$E(k_1,...k_N)$ is the energy of the ground state (GS), i.e., the vacuum of Bogoliubov quasi-particles \cite{21,23,32}, in terms of ${\lbrace}k_j{\rbrace}$ transcribed from Hamiltonian $H$ and the state $|{\psi}_s{\rangle}$. Plugging Eq.~(\ref{eq11}) into Eq.~(\ref{eq8}) and taking derivatives, we find the current contains two terms: \textit{band current} $J_{band}$ and \textit{correlation current} $J_{corr}$
\begin{equation}
\begin{split}
J=q{\langle}\dot{r}{\rangle}_s=J_{band}+J_{corr},
\end{split}
\label{eq13}
\end{equation}
where
\begin{equation}
\begin{split}
J_{band}={\rho}\frac{a}{(2\pi)}\frac{1}{N}{\sum}_j^N{\int}_0^{\frac{2\pi}{a}}\frac{1}{\hbar}{\partial}_{k_j}E(k_1,...k_N)dk_1...dk_N,
\end{split}
\label{eq14}
\end{equation}
and
\begin{equation}
\begin{split}
J_{corr}={\rho}{\cdot}{\partial}_t\frac{a}{2\pi}{\int}_0^{\frac{2\pi}{a}}{\langle}A_k(t)|i{\partial}_kA_k(t){\rangle}dk.
\end{split}
\label{eq15}
\end{equation}

The first term $J_{band}$ involves ${\partial}_kE$, namely the ``band velocity" \cite{19,20} that can be deduced from the velocity operator in Eq.~(\ref{eq1}).  Thus, $J_{band}$ corresponds to the conventionally known superflow in SC (i.e., $j_s$ in London equations). Notably, $\partial_kE$ was derived by applying $\hat{v}$ to Bloch waves \cite{19} $\int_r \psi_{n,k}^*(r)(-i\nabla_r)\psi_{n,k}(r)$. Therefore, it relies on (i) the single-particle approximation, (ii) the crystalline symmetry (i.e., $k$ being a good quantum number). In contrast, deriving Eq.~(\ref{eq14}) merely relies on the fundamental commutator relation and therefore possesses broader applicability. 

The second term $J_{corr}$ integrates the Berry connection yielding a global geometric phase beyond the local operator $\hat{v}$. It formally resembles Eq.~(\ref{eq2}) that characterizes the movement of bound charges. Thus, $J_{corr}$ is intuitively interpreted as the shift of the charge center for a Cooper pair in SC, in analog with the dielectric polarization change for the bound charge in an insulator. Therefore, it requires broken IS and is typically small. However, $J_{corr}$ behaves differently from $j_s$ in terms of the distribution regions, the self-decay tendency, etc., making $J_{corr}$ recognizable even if $J_{band}$ is dominant in magnitude. 

Note that, although Eq.~(\ref{eq10}) is generic, the two terms $J_{band}$ and $J_{corr}$ are specific for intra-band transport of BCS states. In other words, when Eq.~(\ref{eq10}) is applied to different correlations or dynamical processes (e.g., inter-band is included), different or additional terms may emerge. Next, we examine $J_{band}$ and $J_{corr}$ in detail. 

The $J_{band}$ differs subtly from the conventional ``group velocity” current. First, Eq.~(\ref{eq14}) contains $N$ state variables ${\lbrace}k_j{\rbrace}_N$, allowing for the energy to be correlated. Thus, the current formulation remains unchanged with the presence of correlation. We can write Eq.~(\ref{eq14}) as
\begin{equation}
\begin{split}
J_{band}={\rho}\frac{a}{2\pi}{\int}_0^{\frac{2\pi}{a}}\frac{1}{\hbar}{\partial}_k\tilde{E}(k)dk={\rho}\frac{a}{2\pi}{\int}_0^{\frac{2\pi}{a}}\tilde{v}(k)dk
\end{split}
\label{eq16}
\end{equation}
where
\begin{equation}
\begin{split}
\tilde{E}(k):={\int}_0^{\frac{2\pi}{a}}E(k,k_1,...k_{N-1})dk_1...dk_{N-1}.
\end{split}
\label{eq17}
\end{equation}
The $\tilde{E}(k)$ ($\tilde{v}(k)$) means the band energy (velocity) renormalized by correlations (otherwise, $\tilde{E}(k)$ and $\tilde{v}(k)$ reduce to their non-interacting counterparts). Therefore, the connection between ${\partial}_k\tilde{E}(k)$ and velocity has a fundamental quantum origin, making it unnecessary to interpret ${\partial}_k\tilde{E}(k)$ as a ``group velocity". Even when the conditions for a wave-packet description breaks down, the form of ${\partial}_k\tilde{E}(k)$ remains valid for evaluating the velocity and current.

To compute $E(k_1,...k_N)$, we make a progressive count by orders
\begin{equation}
\begin{split}
E(k_1,...k_N)={\sum}_j^NE^{(1)}(k_j)+{\sum}_{j,l}^{\frac{N(N-1)}{2}}E^{(2)}(k_j,k_l)+...
\end{split}
\label{eq18}
\end{equation}
For instance, $E^{(2)}$ is the correlation between two momenta $k_j$ and $k_l$ \cite{33}. However, if free particles or mean-field theory are taken (as in this work), only $E^{(1)}(k_j)$ remains.

A second difference is that the integration in Eq.~(\ref{eq14}) is always over the \textit{entire} BZ, rather than the occupied ${\int}_{occ.}$. The ${\oint}_{BZ}$ ensures gauge invariance and implies that a globally smooth $E(k_1,...k_N )$ must result in $J_{band}=0$. That is because dynamic evolution is similar to U(1) gauge transformation $T_G$ \cite{6,16,27}
\begin{equation}
\begin{split}
T_G:~|{\psi}_s{\rangle}~{\rightarrow}~e^{i{\xi}(k_1,...k_N)}|{\psi}_s{\rangle}.
\end{split}
\label{eq19}
\end{equation}
Equation~(\ref{eq11}) suggests
\begin{equation}
\begin{split}
e^{-iE(k_1,...k_N)t/{\hbar}}=e^{i{\xi}(k_1,...k_N)}
\end{split}
\label{eq20}
\end{equation}
Since Eq.~(\ref{eq10}) respects U(1) gauge symmetry, a smooth $T_G$ cannot alter ${\langle}r{\rangle}$. Only if $e^{-iE(k_1,...k_N)t/{\hbar}}$ is “singular” (i.e., $E(k_1,...k_N)$ involves non-smoothness), can ${\langle}r(t){\rangle}$ move. 

To ``visualize" the singularity in $E(k_1,...k_N)$, we apply the formula Eq.~(\ref{eq13}) derived from BCS state to a special case of ${\Delta}=0$, where the Cooper pairs tend to vanish and the correlated state approaches to a free-electron limit. In logic, we are using a simple case to demonstrate the general mechanism of singularity generating charge motion, to elude unnecessary complexities, such as the singularity source (the presence of Fermi surface, electron-phonon coupling, or others), complicated singular functions for $E(k_1,...k_N)$. Then we move on to the general case ${\Delta}{\neq}0$ to consider a next-tier question: what happens if Fermi surface is gapped and the singularity is provided by electron-phonon coupling instead? Nonetheless, since the free-particle state is an analytic limit of Cooper pair in this framework, the formula remains valid. 

\begin{figure}
\includegraphics[scale=0.27]{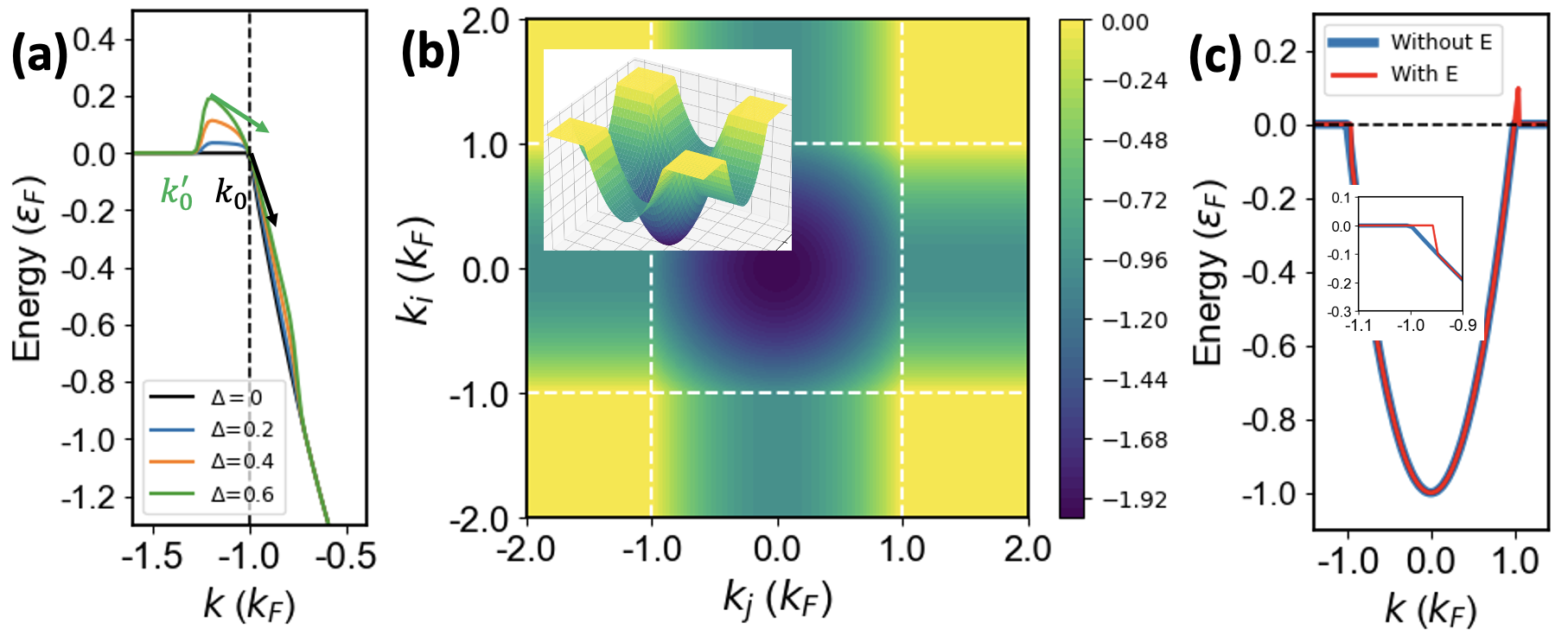}
\caption{\label{fig:epsart}(color online): (a) Singularity in $E^{(1)}(k)$ under varied correlations. $\Delta$ tends to shift singularity $k_0$ away from $k_F$ to $k_0'$, and influence $\bar{v}$ (the arrows). (b) The non-smoothness (white dashed) projected to a 2D subspace of the ${\lbrace}k_j{\rbrace}$ space. (c) $E^{(1)}(k)$ of free particles with/without $E$-field. \label{f1}}
\end{figure}

In the free particle limit, we have
\begin{equation}
\begin{split}
E(k_1,...k_N)={\sum}_j^NE^{(1)}(k_j)
\end{split}
\label{eq21}
\end{equation}
where (derivation in Appendix A)
\begin{equation}
\begin{split}
E^{(1)}(k_j)=
	\begin{cases}
		&2(\frac{{\hbar}^2k_j^2}{2m}-{\epsilon}_F), k_j<k_F \\
		&0, k_j>k_F
	\end{cases}
\end{split}
\label{eq22}
\end{equation}
For free particles, the singularity in $E(k_1,...k_N)$ is due to the non-smooth energy slope at $k_F$ (Fermi energy ${\epsilon}_F:=({\hbar}^2k_F^2)/2m$), which happens to border the occupied and empty subspaces. (Fig.~\ref{f1}a, b). This leads to a decomposition ${\int}_{k<k_F}+{\int}_{k>k_F}$, which reduces Eq.~(\ref{eq14}) to the familiar sum rule for occupied velocities ${\int}_{occ.}\tilde{v}(k)dk$ (Appendix). For BCS states, the singular $k_0$ is different from $k_F$. The original singularity at $k_F$ is smoothed out by SC gap, while $k_0'$ emerges at the edge of Debye energy $k_B{\Theta}$ (Fig.~\ref{f1}a). 

Physically, the non-smoothness in $E(k_1,...k_N)$ of BCS states is due to electron-phonon coupling. Consequently, integration concerns a margin ${\sim}k_B{\Theta}$ above $k_F$, at which electron's coupling with optical phonons abruptly disappears. 
Although the current formulation involves no phonon wavefunction, phonon quantization ($k_B\Theta$) affects the singularity in ${\Delta}(k)$ function, serving a similar role as Fermi surface in free particle.

Thus, the familiar sum $\int_{occ}$ is reproduced by a geometric method, which is more general, because it works for states without Fermi surface or a clear boundary between the occupied and empty regions. Next, consider a more general case ${\Delta}{\neq}0$. The net velocity $\bar{v}:=\frac{a}{2\pi}{\int}_0^{\frac{2\pi}{a}}\tilde{v}(k)dk$ is most related to $\tilde{v}(k)$ at $k_0$: $\bar{v}{\approx}\tilde{v}(k_0)$. In Fig.~\ref{f1}a, we show how $\Delta$ shifts the singular points and modifies $\bar{v}$. 

Regarding the physical significance of $J_{band}$, it is the microscopic correspondence for $j_s$. Given $\Delta{\neq}0$,
\begin{equation}
\begin{split}
E^{(1)}(k_j)={\epsilon}_{k_j}-\frac{{\epsilon}_{k_j}^2}{\sqrt{{\epsilon}_{k_j}^2+{\Delta}_{k_j}^2}}-\frac{{\Delta}^2}{NV},
\end{split}
\label{eq23}
\end{equation}
replaces $E^{(1)}(k_j)$ in Eq.~(\ref{eq22}) for free particles. Consider a static SC with constant $\Delta$. Then ${\partial}_t|A_k(t){\rangle}=0$ and $J_{corr}=0$. Eq.~(\ref{eq10}),(\ref{eq11}) give ${\langle}r(t){\rangle}=t{\cdot}\bar{v}$, ($\bar{v}$ is determined by ${\epsilon}_k$ and $\Delta$). That means, charges in BCS states tend to move with constant velocity unless they are disrupted. The non-decay supercurrent can be attributed to the persistent movement from $J_{band}$. Note that such movement is deduced form the pure GS, not from quasi-particles \cite{35,36,37}. It is also different from understanding transports as hopping between eigenstates in a perturbed Hamiltonian \cite{9,38,39}.

To make $\bar{v}{\neq}0$, we need to break IS, e.g., by applying an $E$-field, which makes the total energy $E(k_1,...k_N)$ asymmetric (Fig.~ \ref{f1}c) by shifting the electronic state slightly away from GS (Appendix). According to Eq.~(\ref{eq14}), $E(k_1,...k_N)$ determines $J_{band}$. Intuitively, the non-decay current is due to acceleration of momentum to make ${\langle}p{\rangle}{\neq}0$. However, we will see there is a self-decay current component $J_{corr}$ in SC and when ${\langle}p{\rangle}=0$, it is possible $J{\neq}0$. 

The third crucial difference is that, in Eqs.~(\ref{eq16}), only the integral of $\tilde{v}(k)$ is an observable \cite{34}. By contrast, conventionally, every local value $v(k)$ is regarded as an observable, representing the velocity of a wave packet with momentum $k$. Intuitively, an extended Bloch wave is interpreted as an array of ``bumps", each representing a classical-like particle, arranged periodically in space. Every packet is assigned a definite position (the packet center) and an overall velocity (the group velocity). Assigning $v$ to a packet instead of a fundamental electron serves to circumvent the difficulty from uncertainty principle \cite{19,20}. However, an evident contradiction is that a stable physical object (electron's lifetime $>10^{26}$ years) is depicted by a unstable construct, since a wave packet readily disperses due to $k$-dispersion. 

The geometric formula provides alternative paths for reconciling with the uncertainty principle: only the holistic velocity is a physical observable, removing the necessity of chopping the wave into ``bumps". In other words, it is meaningful to discuss the net velocity for all the bumps, but not for each individual.

Regarding the second term $J_{corr}$, it suggests that, even with $J_{band}=0$, i.e., the net momentum is zero, the time-dependent correlation is still able to cause charge movements. While existing methods normally assign $k$ with independent velocity \cite{11,12,13} (which encountered surface terms or uncontrollable errors \cite{40,41,42,43,44}), $J_{corr}$ suggests that correlation among $k$ (e.g., $\Delta(t):={\langle}c_kc_{-k}{\rangle}$) gives extra contributions. 





\begin{figure}
\includegraphics[scale=0.32]{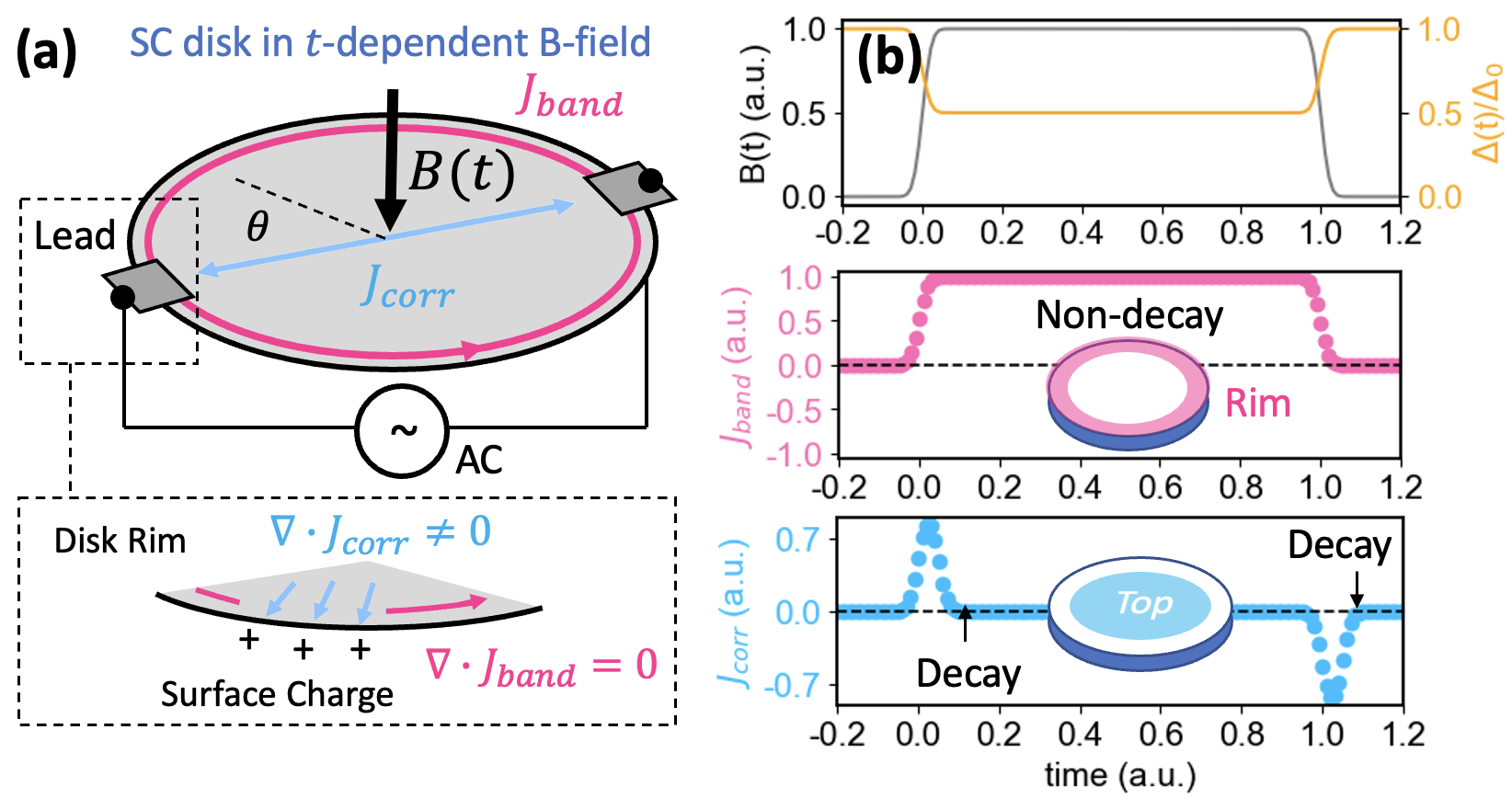}
\caption{\label{fig:epsart}(color online): (a) Scenarios for $J_{corr}$ along the IS breaking axis, away which $J_{corr}$ will decay (b) Time dependence and space distribution for $J_{band}$, $J_{corr}$ under a perturbation $\boldsymbol{B}(t)$. \label{f2}}
\end{figure}

To find $J_{corr}$, we examine a generic $|{\psi}_s{\rangle}={\prod}_j(u_{k_j}+e^{i{\phi}(k_j)}v_{k_j}b_{k_j}^{\dagger})|{\Phi}_0{\rangle}$ that allows complex coefficients. Then,
\begin{equation}
\begin{split}
J_{corr}=-\frac{a}{2\pi}{\oint}\frac{{\epsilon}(k){\Delta}(k){\phi}'(k)\dot{\Delta}_k}{(\sqrt{{\epsilon}_k^2+{\Delta}_k^2})^3}dk.
\end{split}
\label{eq24}
\end{equation}
To make $J_{corr}{\neq}0$, we need to break IS, for instance, by choosing a non-centrosymmetric SC with ${\epsilon}_k{\neq}{\epsilon}_{-k}$ \cite{45,46}, such as hetero-structures \cite{47,48}, FeSe \cite{49}. 
$\dot{\Delta}_k$ tunes the correlation strength to transition the phase \cite{50,51,52,53}, such as applying magnetic field (Fig.~\ref{f2}a), or by controlling potentials in an optical lattice \cite{54,55,56}. For simplicity, we consider a type-I SC disk, where $\boldsymbol{B}(t)$ is along $z$ to render ${\Delta}_k$ time-dependent within the penetration depth. The $J_{corr}$ will be induced by $\boldsymbol{B}(t)$ along the IS breaking axis. 

The detection of $J_{corr}$ is facilitated by the presence of a boundary, where ${\nabla}{\cdot}J_{corr}\neq 0$ leads to charge accumulation on the surface (Fig.~\ref{f2}a), and consequently to a voltage difference -- an effect that is anomalous from the conventional picture of SC as an equal-potential system. Moreover, $J_{corr}$ reverses direction under a periodic perturbation $\boldsymbol{B}(t)$ (Fig.~\ref{f2}b), producing oscillatory signals, whereas $J_{band}{\propto}|\boldsymbol{B}|$ does not exhibit such behavior.

A recent experimental breakthrough involves the light-induced IS breaking in superconductors \cite{1a,2a}. With Terahertz (THz) pulses, we induce supercurrents that dynamically distribute Cooper pairs with finite center-of-mass momenta, persisting significantly beyond the pulse duration \cite{3a,4a}. These findings offer compelling insights into exploring band structures and $J_{corr}$ under nonlinear driven conditions in various superconductors \cite{5a,6a,7a}.

The $J_{corr}$ should not be confused with the supercurrent $j_s$ in London equations.\cite{5} First, $J_{corr}$ in Fig.~\ref{f2}a distributes near the top of the disk, while $j_s$ circulates around the rim and generates a compensating magnetic field inside SC (i.e., Meisner effect). Second, $J_{corr}$ is driven by correlation change, while $j_s$ is accelerated by induced field $\boldsymbol{E}=-{\partial}_t\boldsymbol{B}$. 
Third, $J_{corr}$ vanishes once the driving ceases, i.e., when $\dot{\boldsymbol{B}}(t)=0$, whereas $j_s$ arising from $J_{band}$ should persist (see the comparison in Fig.~\ref{f2}b). On the other hand, if IS is preserved or if the correlations $u_k$, $v_k$ are real, $J_{corr}=0$, and only $J_{band}$ remains. In this limit, the current reduces to the conventional momentum-based expressions \cite{2,13}.

\section{IV. Discussion} 
Essentially, the present formulation extends the previous phase-based formulation \cite{24,25,26} from a vector space spanned by single-particle states to one encompassing varying particle numbers, as implemented through Eq.~(\ref{eq5}) together with the procedures in Eqs.~(\ref{eq4})(\ref{eq7}). In this framework, the particle number no longer plays a central role in defining $\langle r \rangle$ or $J$. By contrast, the operator $\hat{v}=-i\hbar\sum_j\nabla_{r_j}$ intrinsically depends on the particle number, since the coordinates $r_j$ are tied to the particle number, rendering the formulation cumbersome once particles are allowed to be created or annihilated.

States with different particle numbers are uniformly represented by vectors in the second-quantized Hilbert space, and contribute on equal footing when the state vector is twisted to generate the Berry phase. Even the vacuum state (corresponding in the BCS state to $u_{k_j}=1$ and $v_{k_j}=0$) can contribute to $\vartheta$ and hence to $J$ when it appears as a component of the correlated state. Strictly speaking, the contribution by the vacuum state cannot be defined without specifying the the global structure of the many-body state. By contrast, the operator $\hat{v}$ simply suggests a vanishing contribution from the vacuum state. 

The phase-based formulation employs $\hat{r}$ instead of $\hat{v}$ to define the velocity or current density. The key distinction is that the form of $\hat{v}=(1/i\hbar)[\hat{r},H]$ depends explicitly on the Hamiltonian $H$, whereas $\hat{r}$ retains the ``universal form" Eq.~(\ref{eq4}), being determined solely by the fundamental commutation relationship -- the $H$ influences $\langle r \rangle$ through affecting the wavefunction rather than modifying the expression of operator. In addition, $\hat{v}$ is tied to local $k$ \cite{11,12,13}, while $\langle \hat{r} \rangle$ is equal to a ``global" Berry phase. Since the Berry phase is intrinsically U(1) gauge invariant, the proposed current formulation naturally yields a gauge-invariant current from the gauge-variant BCS state, as generally required for current \cite{16,17,18}. 




Here, geometry accounts for the charge center $\langle r \rangle$ of the many-body wavefunction, instead of formulating the transport coefficients, as the case in quantum Hall effect (QHE). Then, should $\langle r \rangle$ be quantized by the geometric formulation? If it were, it would imply that charges are forbidden from occupying certain spatial regions or that spatial motion occurs only through hopping between discrete points, both of which are physically implausible.

Regarding quantization, QHE arises from integrating the Berry curvature over a closed manifold \cite{c11,c12,c23}. In contrast, integrating a general geometric quantity, such as the metric, does not necessarily lead to quantization. \cite{c3,c4,c5} In the present case, it integrates the Berry connection, yielding a continuous-valued Berry phase that represents $\langle r \rangle$.

The QHE involves the microscopic wavefunction and quantization (characterizing the topological states). The flat-band SC shows that geometry (quantum metric) may take effect even without quantization, while its geometry is defined on the free energy instead of the microscopic wave function. In the present case, geometry enters via operator $\hat{r}$, not limited to a specific microscopic state or quantization.

The Berry-phase formula reveals SC’s dependence on geometric aspects of quantum \cite{22,24,58,59,60,61,62,63}, and it is possible to be extended to other SCs \cite{47,48,49,64,65,66,67} or correlated states \cite{68,69,70}. More broadly, Berry phase has previously been applied to constructing plausible many-body wavefunctions for the fractional quantum Hall effect (FQHE) \cite{71}. A central challenge in understanding the FQHE, addressed by the Berry phase or other proposals \cite{71,72,73}, is the emergence of fractional filling subject to interaction. In that context, the Berry-phase framework aims to identify the microscopic states responsible for the fractional quantization. For the superconducting current formula studied here, the microscopic wave function is already established. The key questions instead are: first, whether the current formulation is complete for a given state; second, whether the SC current takes a unified form with non-SC. Therefore, in the two contexts, Berry phases address questions at fundamentally different levels.

\section{V. Conclusion}
Equations~(\ref{eq13})-(\ref{eq15}) represent generic ways of calculating position center ${\langle}r{\rangle}$ and current ${\langle}\dot{r}{\rangle}$ for extensive wavefunctions that involve correlation and particle number variation. Applied to SC, the current contains $J_{band}$ and $J_{corr}$, both arising from GS condensates rather than quasi-particle excitation. The supercurrent is attributed to $J_{band}$, taking a form similar to group velocity. By contrast, $J_{corr}$ emerges when correlation is time dependent and IS is broken. The distinguishable features of $J_{corr}$ relative to $J_{band}$ make it observable in time-dependent $B$-field, optical lattices \cite{54,55,56}. Inspiring perspectives emerge, such as connections between current generation and singularity in energy $E(k_1,...k_N)$.

\input acknowledgement.tex   

\appendix

\section{Appendix A: Derive $E(k_1,...,k_N)$ at $\Delta=0$ and $\Delta\neq 0$.}
The ground state energy of the BCS Hamiltonian is $E(k_1,...k_N)$, defined by Eq.~(\ref{eq12}). To find it, we plug the GS wavefunction Eq.~(\ref{eq3}) into the BCS Hamiltonian
\begin{equation}
\begin{split}
H={\sum}_{k_j,{\sigma}}c_{k_j,{\sigma}}^{\dagger}c_{k_j,{\sigma}}+{\sum}_{k_j,k_l}V_{k_j,k_l}c_{k_j,{\uparrow}}^{\dagger}c_{-k_j,{\downarrow}}^{\dagger}c_{-k_l,{\downarrow}}c_{k_l,{\uparrow}}
\end{split}
\label{eq25}
\end{equation}
with $V_{k_j,k_l}$ as an attractive potential (negative). For simplicity, $V_{k_j,k_l}$ is assumed vanishing at Debye energy $\hbar{\omega}_D$:
\begin{equation}
\begin{split}
V_{k_j,k_l}=
	\begin{cases}
		&-V,~|{\epsilon}_{k_j}|,|{\epsilon}_{k_l}|<\hbar{\omega}_D;  \\
		&~0,~~\text{otherwise.}
	\end{cases}
\end{split}
\label{eq26}
\end{equation}
Then ${\langle}{\psi}_s|H|{\psi}_s{\rangle}=$
\begin{equation}
\begin{split}
{\sum}_j2v_{k_j}^2{\epsilon}_{k_j}+{\sum}_{k_j,k_l}V_{k_j,k_l}u_{k_j}v_{k_j}u_{k_l}v_{k_l}.
\end{split}
\label{eq27}
\end{equation}
Introducing ${\Delta}_{k_j}:=-{\sum}_lV_{k_j,k_l}u_{k_l}v_{k_l}$
and combining with Eq.~(\ref{eq26}), we find 
\begin{equation}
\begin{split}
{\Delta}_{k_j}=
	\begin{cases}
		&{\Delta},~|{\epsilon}_{k_j}|<\hbar{\omega}_D;  \\
		&~0, |{\epsilon}_{k_j}|>\hbar{\omega}_D.
	\end{cases}
\end{split}
\label{eq28}
\end{equation}
Then Eq.~(\ref{eq27}) becomes $E(k_1,...k_N)=$
\begin{equation}
\begin{split}
{\sum}_j^N\left({\epsilon}_{k_j}-\frac{{\epsilon}_{k_j}^2}{\sqrt{{\epsilon}_{k_j}^2+{\Delta}_{k_j}^2}}-\frac{{\Delta}^2}{NV}\right).
\end{split}
\label{eq29}
\end{equation}
Given the first order truncation $E(k_1,...k_N)={\sum}_j^NE^{(1)}(k_j)$, based on Eq.~(\ref{eq29}), clearly Eq.~(\ref{eq23}) gives $E^{(1)}(k_j)$ for BCS states. For free particles, we simply set ${\Delta}=0$, and Eq.~(\ref{eq29}) becomes
\begin{equation}
\begin{split}
E(k_1,...k_N)={\sum}_j^N({\epsilon}_{k_j}-|{\epsilon}_{k_j}|).
\end{split}
\label{eq30}
\end{equation}
For $k_j<k_F$, ${\epsilon}_{k_j}<0$, and $E(k_1,...k_N)={\sum}_j^N2{\epsilon}_{k_j}$; for $k_j>k_F$, ${\epsilon}_{k_j}>0$, $E(k_1,...k_N)=0$. This is exactly Eq.~(\ref{eq22}) in the main text.

The GS energy is the lowest energy under the Hamiltonian, which is a single value. Now $E(k_1,...k_N)$ is not a value, but a function of $N$ variables. Clearly, when $k_j=j{\cdot} 2{\pi}/Na$ is substituted, $E(k_1,...k_N)$ is the GS energy. What does $E(k_1,...k_N)$ mean when ${\lbrace}k_j{\rbrace}$ takes other values? It is tempting to think that $E(k_1,...k_N)$ might mean the excitation energy when $k_j{\neq}j{\cdot}2{\pi}/Na$. However, this interpretation is incorrect. 

To clarify, we write the full variable labels of ${\langle}{\psi}_s|H|{\psi}_s{\rangle}$:
\begin{equation}
\begin{split}
{\langle}{\psi}_s(k_1,...k_N)|H(k_1,...k_N)|{\psi}_s(k_1,...k_N){\rangle}.
\end{split}
\label{eq31}
\end{equation}
Not only the state vector $|{\psi}_s{\rangle}$ is characterized by ${\lbrace}k_j{\rbrace}$: the Hamiltonian $H$ is also a function of ${\lbrace}k_j{\rbrace}$. Thus, $E(k_1,...k_N)$ is understood as the GS energy for the parameterized Hamiltonian $H(k_1,...k_N)$.

It would be incorrect to consider $E(k_1,...k_N)={\sum}_j^N(\frac{{\hbar}^2k_j^2}{2m}-{\epsilon}_F) $, without distinguishing a range below or above ${\epsilon}_F$. In that case, $E(k_1,...k_N){\neq}{\lim}_{{\Delta}{\to}0}{\langle}{\psi}_s|H|{\psi}_s{\rangle}$, and it will mean the excitation energy when $k_j{\neq}j{\cdot}2{\pi}/Na$. Here, $E(k_1,...k_N)$ appears in the dynamic phase of GS. Thus, it must be defined with Eq.~(\ref{eq12}), meaning the total energy of GS.

Evidently, $E(k_1,...k_N)$ is not smooth with respect to coordinates ${\lbrace}k_j{\rbrace}$. The non-smoothness is easy to imagine for free particles, as shown by Eq.~(\ref{eq22}) and Fig.~\ref{f1}a, b. Below ${\epsilon}_F$, the energy is parabolic; above ${\epsilon}_F$, the energy is flat, since no state is occupied. 

It is important to remember that the singularity refers to the total energy, rather than the energy of an individual quasi-particle, although for free particles the two viewpoints could be reconciled. The gauge transformation Eq.~(\ref{eq19}) and dynamic evolution Eq.~(\ref{eq11}) are also associated with a single vector for a multiple-particle state. It would be misleading to imagine $N$ independent evolutions for each single particle. In general, the non-smoothness $E(k_1,...k_N)$ is not restricted to free particles. Importantly, for correlated wavefunctions, $k_0{\neq}k_F$ is possible. The non-smoothness is shown in Fig.~\ref{f1} for both free particles and BCS states. We denote the singular point with $k_0$, at which the slope is divergent.

To perform integration with isolated singular points $k_0$, we simply need to integration on each side of $k_0$. For free particles, we obtain
\begin{equation}
\begin{split}
\frac{a}{2\pi}{\int}_{-\frac{\pi}{a}}^{\frac{\pi}{a}}\tilde{v}(k)dk={\int}_{-\frac{\pi}{a}}^{-k_F}+{\int}_{-k_F}^{+k_F}+{\int}_{+k_F}^{\frac{\pi}{a}}.
\end{split}
\label{eq32}
\end{equation}
Clearly, $\tilde{v}(k)\equiv0$ in ${\int}_{-\frac{\pi}{a}}^{-k_F}$ and ${\int}_{k_F}^{\frac{\pi}{a}}$. Thus, we only need to count ${\int}_{-k_F}^{k_F}$, which is exactly the occupied region in the BZ. It is convenient to write the continuous integration in discrete form to demonstrate the quasi-particles: 
\begin{equation}
\begin{split}
\frac{a}{2\pi}{\int}_{-\frac{\pi}{a}}^{\frac{\pi}{a}}\tilde{v}(k)dk={\sum}_{k_j}^{occ}\tilde{v}(k_j).
\end{split}
\label{eq33}
\end{equation}
However, for BCS states (${\Delta}{\neq}0$), $k_0{\neq}k_F$. Thus Eq.~(\ref{eq34}) is unachievable: The summation must be extended beyond the states that were originally empty. For free particles, there is a clear-cut boundary between the occupied and empty regions. However, when the Fermi surface is gapped, the boundary becomes vague. Thus, the summation of velocity for occupied states is only rigorously true for free particles.

Next, we try to find the total system energy $E'(k_1,...k_N)$ under an electric field. The key idea is that the state perturbed by an $E$-field can be characterized with asymmetric Fermi energies, and $E'(k_1,...k_N)$ is evaluated by
\begin{equation}
\begin{split}
E'(k_1,...k_N)={\langle}{\psi}'_s|H|{\psi}'_s{\rangle}.
\end{split}
\label{eq34}
\end{equation}
The perturbed $|{\psi}'_s{\rangle}$ can be expressed by the generic wave function Eq.~(\ref{eq5}) with an updated $u'_k$ and $v'_k$. For the GS, there exists a common $k_F$ for the $\pm k$ 
directions; with the $E$-field, we introduce $k_F^{L}$ and $k_F^R$ to describe the asymmetric wave function that breaks IS. 

We find $\frac{E^{(1)'}(\kappa)}{{\epsilon}_F}+\frac{{\Delta}^2}{NV{\epsilon}_k}$ is equal to
\begin{equation}
\begin{split}
({\kappa}^2-1)-\text{sgn}(({\kappa}-1)({\kappa}-{\kappa}_0^X))\frac{({\kappa}^2-1)^2}{\sqrt{({\kappa}^2-1)^2+{\Delta}_k^2}}.
\end{split}
\label{eq35}
\end{equation}
with $X=R$ for $k>0$ and $X=L$ for $k<0$,
where ${\kappa}:=k/k_F$ and ${\epsilon}_F$ is the unperturbed Fermi energy. In the absence of electric field, we have $k_0^R=k_0^R=k_F$, i.e., ${\kappa}_0^L={\kappa}_0^R=1$, such that Eq.~(\ref{eq36})
will reduce to Eq.~(\ref{eq23}). For the free particles ${\Delta}=0$, as plotted in Fig.~\ref{f2}b.

\section{Appendix B: Crossover to Normal Metals and Insulators.}
In this section, we show the forms of current in two normal (non-interacting) phases: metals and insulators. The geometric formulation will tend to the known forms of currents: the metallic and insulator limits correspond to the cases of $J_{corr}=0$ and $J_{band}=0$, respectively.

First, consider a single metallic band in a normal state. Clearly, $\Delta(k)\equiv 0$ vanishes the integral Eq.~(\ref{eq24}), making $J_{corr}=0$. For $J_{band}$, plugging the energy Eq.~(\ref{eq22}) into the current formula Eq.~(\ref{eq16}), we have
\begin{equation}
\begin{split}
&J_{band}=\rho\frac{a}{2\pi}\frac{1}{N}\sum_j\int_{|k|<k_F} \frac{1}{\hbar}{\partial}_{k_j}(2\epsilon_{k_j}-2\epsilon_F){\cdot}dk_j\\
&=\rho\frac{a}{2\pi}\int \frac{1}{\hbar}{\partial}_{k}(2\epsilon_{k}){\cdot}dk=\rho\frac{a}{2\pi}\sum_s\int \frac{1}{\hbar}\frac{\partial\epsilon_k}{\partial k}{\cdot}dk,
\end{split}
\label{eq36}
\end{equation}
where $\epsilon_k$ is the normal-state band energy. Thus, the geometric formula reduces to a current for a partially filled band with spin degeneracy. 

Second, consider an insulator. Clearly, $J_{band}=0$ because $\int\to\oint$. The motion of bound charge in insulators is due to changes of electric polarization, and a non-trivial model should contain at least two bands. Thus, we extend the one-band BCS model Eq.~(\ref{eq5}) into the following
\begin{equation}
\begin{split}
\begin{pmatrix} a_{k_1}u_{k_1} \\ a_{k_1}v_{k_1} \\ b_{k_1}u_{k_1} \\ b_{k_1}v_{k_1} \end{pmatrix}{\otimes}...{\otimes}\begin{pmatrix} a_{k_N}u_{k_N} \\ a_{k_N}v_{k_N} \\ b_{k_N}u_{k_N} \\ b_{k_N}v_{k_N} \end{pmatrix} \to \Pi_j\begin{pmatrix} 0 \\ a_{k_j} \\ 0 \\ b_{k_N} \end{pmatrix}.
\end{split}
\label{eq37}
\end{equation}
The coefficients $a_k$ and $b_k$ are associated with two orbital bands, and the previous one-band model corresponds to $a_k \equiv 1$ and $b_k \equiv 0$. With the two-band model, ``twisting" is allowed to yield a non-zero Berry phase.

In a fully filled band with $\Delta=0$, we have $u_{k_j}\equiv0$ and $v_{k_j}\equiv 1$ for all $k_j$ in BZ. Then the state vector reduces to a two-component form as indicated by Eq.~(\ref{eq37}). Plug the state vector into Eq.~(\ref{eq15}), we find
\begin{equation}
\begin{split}
J_{corr}\to J_{ad}=\frac{-e}{2\pi}\partial_t\oint \begin{pmatrix} a_k^* & b_k^* \end{pmatrix} \partial_k \begin{pmatrix} a_k \\ b_k \end{pmatrix} dk
\end{split}
\label{eq38}
\end{equation}
The $J_{corr}$ crossovers to the adiabatic current associated with orbital electric polarization characterized by the vector $(a_k,b_k)^T$. Thus, the adiabatic current and correlation current have a common microscopic origin and both require IS breaking.




\end{document}

%% file: acknowledgement.tex
\textbf{Acknowledgement}. We wish to acknowledge the inspiring discussion about experimental observation and other aspects with David C. Johnston, Pavan Hosur in the course of preparing this work. 

This work was supported by the Ames National Laboratory, the US Department of Energy, Office of Science,
Basic Energy Sciences, Materials Science and Engineering Division under contract No. DEAC02-07CH11358. This work was supported by the Department of Energy grant number DE-SC0022264.

%% file: SCCurrent_5.bbl
\begin{thebibliography}{99}
\bibitem{29} The ${\langle}r{\rangle}$ is now equal to Berry phase $\vartheta$, which is well-known to be gauge invariant.
\bibitem{28} The infinite integration ${\int}_{-{\infty}}^{\infty}dr$ has become a finite ${\int}_{-{\infty}}^{\infty}dk$ in Eq.~\ref{eq2}, leading to a convergent result.
\bibitem{30} In general, isomorphism is an reversible structure-preserving map. Here, isomorphism means the one-to-one map: ${\prod}_j^N(u_{k_j}+v_{k_j}b_{k_j}^{\dagger})|{\Phi}_0{\rangle}{\mapsto}\begin{pmatrix} u_{k_1} \\ v_{k_1} \end{pmatrix}{\otimes}...\begin{pmatrix} u_{k_N} \\ v_{k_N} \end{pmatrix}$. If we make such a substitution indicated by the map, we should obtain the same inner product as expressed by Eq.~\ref{eq6}.
\bibitem{33} Note that $k_j$, $k_l$ are not associated with particular particles, such as the $j^{th}$, $l^{th}$ particles. It should be comprehended as inter-state correlation.
\bibitem{34} $\tilde{v}(k)$ depends on energy, thus it is gauge invariant. However, gauge-invariance is only a necessary condition for an observable.
\end{thebibliography}
